# An LLM-Enabled Frequency-Aware Flow Diffusion Model for Natural-Language-Guided Power System Scenario Generation

Zhenghao Zhou, *Student Member, IEEE*, Yiyan Li, *Member, IEEE*, Fei Xie, Lu Wang, Bo Wang, Jiansheng Wang, Zheng Yan, *Senior Member, IEEE*, and Mo-Yuen Chow, *Fellow, IEEE*

*Abstract*—Diverse and controllable scenario generation (e.g., wind, solar, load, etc.) is critical for robust power system planning and operation. As AI-based scenario generation methods are becoming the mainstream, existing methods (e.g., Conditional Generative Adversarial Nets) mainly rely on a fixed-length numerical conditioning vector to control the generation results, facing challenges in user conveniency and generation flexibility. In this paper, a natural-language-guided scenario generation framework, named LLM-enabled Frequency-aware Flow Diffusion (LFFD), is proposed to enable users to generate desired scenarios using plain human language. First, a pretrained LLM module is introduced to convert generation requests described by unstructured natural languages into ordered semantic space. Second, instead of using standard diffusion models, a flow diffusion model employing a rectified flow matching objective is introduced to achieve efficient and high-quality scenario generation, taking the LLM output as the model input. During the model training process, a frequency-aware multi-objective optimization algorithm is introduced to mitigate the frequency-bias issue. Meanwhile, a dual-agent framework is designed to create text-scenario training sample pairs as well as to standardize semantic evaluation. Experiments based on large-scale photovoltaic and load datasets demonstrate the effectiveness of the proposed method.

## I. Introduction

THE increasing uncertainty in modern power systems necessitates extensive scenario analysis (i.e. wind, solar, power load, etc.) to ensure reliability in system planning [1] and operation [2]. However, relying solely on scenarios derived from historical data is usually insufficient, as they often fail to capture rare extreme scenarios or hypothetical future operating conditions due to limited observation windows. As such, scenario generation becomes an important solution.

Approaches to scenario generation can be categorized into three primary paradigms: physics-based simulation, statistical modeling, and machine learning methods. Physics-based simulation offers an intuitive approach by modeling the physical mechanisms of real-world scenarios [3]. While it yields interpretable data with direct physical correspondence, it struggles to capture diverse stochastic variations and suffers from low computation efficiency and heavy labor burdens. Statistical methods, such as Latin hypercube sampling, Monte Carlo methods, and Copula models, attempt to replicate scenario distributions [4]. However, these methods often impose strong assumptions on marginal distributions or dependency structures and face scalability issues in high-dimensional settings due to the curse of dimensionality.

With the evolution of artificial intelligence, machine learning methods, particularly deep generative models, have demonstrated capabilities in capturing complex data structures and generating high-quality scenarios. They do not rely on rigid parametric assumptions about the underlying data distribution, instead learning flexible representations directly from data. Variational Autoencoder (VAE) learns temporal latent representations via an encoder-decoder architecture [5]. However, VAE tends to produce blurred samples that lack the sharp details required for realistic power curves. Generative Adversarial Network (GAN) [6], through its adversarial training dynamic, achieves higher visual fidelity but is blamed for issues such as training instability and mode collapse, limiting its their reliability in generating diverse critical scenarios [7], [8].

Recently, Denoising Diffusion Probabilistic Model (DDPM) [9] has advanced the scenario generation by modeling the distribution through a gradual denoising process. Promising frameworks such as DiffCharge [10], PrivLoad [11], EnergyDiff [12], and PDM [13] have demonstrated the ability of diffusion models to synthesize high-fidelity and diversified scenarios. Despite these advancements, applying standard DDPM to scenario generation faces two bottlenecks. First, the iterative solution of stochastic differential equations results in slow inference speeds [14]. Second, existing diffusion objectives are primarily optimized in the time domain, causing a "spectral bias" where the model prioritizes low-frequency trends while over-smoothing high-frequency transients [15]. In power systems, preserving these high-frequency components is critical, as they reflect load volatility and renewable intermittency. Their loss compromises signal

This work was supported by National Natural Science Foundation of China under Grant 52307121 and Grant U24B6009, and also supported by State Power Investment Corporation (SPIC) Class B1 Key Project: "Grid-Friendly Renewable Energy Power Station". (Corresponding author: Yiyan Li.)

Zhenghao Zhou and Yiyan Li are with the College of Smart Energy, Shanghai Non-Carbon Energy Conversion and Utilization Institute, and Key Laboratory of Control of Power Transmission and Conversion, Ministry of Education, Shanghai Jiao Tong University, Shanghai, 200240, China. (e-mail: zhenghao.zhou@sjtu.edu.cn, yiyan.li@sjtu.edu.cn).

Fei Xie and Lu Wang are with Qinghai Photovoltaic Industry Innovation Center Co., Ltd., State Power Investment Corporation, Xining 810007, China. (e-mail: xieflyj@yeah.net, wanglu2202@outlook.com)

Bo Wang and Jiansheng Wang are with Xinjiang Hami Co., Ltd., State Power Investment Corporation, Hami 839000, China. (e-mail: 1012369501@qq.com, 15688327162@163.com)

Zheng Yan is with the Key Laboratory of Control of Power Transmission and Conversion, Ministry of Education, and the Shanghai Non-Carbon Energy Conversion and Utilization Institute, Shanghai Jiao Tong University, Shanghai 200240, China. (e-mail: yanz@sjtu.edu.cn)

Mo-Yuen Chow is with the Global College, Shanghai Jiao Tong University, Shanghai, 200240, China. (email: moyuen.chow@sjtu.edu.cn)

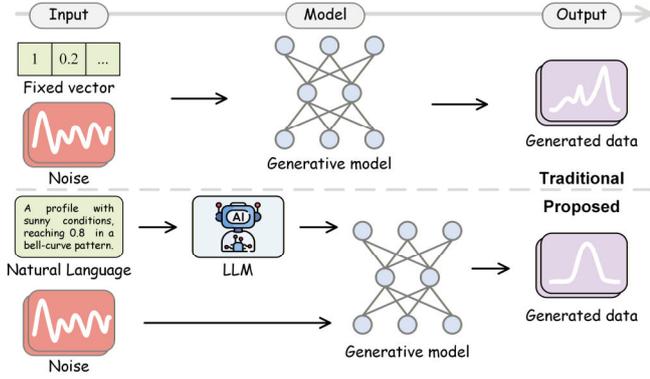

Fig. 1. Illustration about differences between traditional generative models (upper part) and the proposed natural-language-guided model (lower part).

fidelity and reduces the practical utility of the scenario generation.

While fidelity and efficiency are essential, practical application also requires the controllable generation of specific scenarios. To achieve controllable generation, researchers have integrated various conditions into generative models, such as extreme-value quantiles [16], user attributes [17], point forecasts [18], weather dynamics [19], pattern features [20], and socio-demographic attributes [21]. However, these mechanisms usually rely on a fixed-dimension vector as the condition representation, leading to the following two limitations: 1) Users have to specify the values of all positions of the predefined condition vector, even if some positions are irrelevant or hard to acquire, which is user-unfriendly and will introduce noises. 2) Users only have limited generation flexibility by alternating the values of the predefined condition vector, impairing the diversity of the generation results [22].

As illustrated in Fig. 1, to address the above challenges (i.e., the efficiency and frequency-bias issues of DDPM, and the flexibility issue caused by the fixed condition vector), this paper proposes a novel LLM-guided frequency-aware flow diffusion model to achieve flexible, efficient and high-fidelity scenario generation. Large Language Model (LLM) is a powerful tool that can achieve the transformation from unstructured natural languages into quantitative features [23], [24]. Considering this, a pretrained LLM module is introduced to convert the generation requirements described by natural languages form users into structured condition vectors, enabling extensive control flexibility. Then taking the converted condition vector as the input, a flow diffusion model is proposed to achieve efficient, high-fidelity, and customizable scenario generation.

Contributions are summarized as follows:

1. A natural-language-guided generative framework is proposed by combining LLM and a flow diffusion model, enabling scenario generation with high efficiency, flexibility and fidelity.

2. A text-oriented temporal denoising network is designed featuring multi-resolution cross-modal attention, which bridges the gap between unstructured natural language and quantitative scenario characteristics.

3. A frequency-aware multi-objective optimization strategy is proposed by embedding a specially-designed spectral consistency loss into the Multiple Gradient Descent Algorithm (MGDA), ensuring the model can capture both scenario trends and high-frequency transients.

4. A dual-agent framework is designed to create text-scenario training sample pairs as well as to achieve semantic evaluation, which can serve as a standardized pipeline for solving other text-guided tasks in the power domain.

The rest of this paper is organized as follows: Section II introduces the methodology. Section III demonstrates the case study results, and Section IV concludes the paper.

## II. METHODOLOGY

The LLM-enabled Frequency-aware Flow Diffusion (LFFD) framework is proposed to tackle the synthesis problem $p(\mathbf{x}|\mathbf{c})$ by learning a straight-line probability flow guided by a natural language prompt $\mathbf{c}$. The LFFD architecture comprises three tightly coupled components: a text-guided rectified flow matching objective, a frozen LLM for semantic encoding, and a novel denoising network as the velocity field predictor $\mathbf{v}_\theta$. Meanwhile, two LLM-agents are introduced for prompt synthesis and consistency evaluation.

### A. Frequency-Aware Rectified Flow Matching Objective

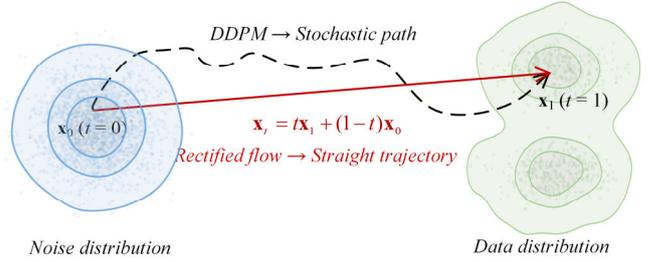

Fig. 2. Rectified flow achieves fast generation via a straight-line trajectory.

Inspired by recent advancements, this work adopts the flow matching framework, specifically utilizing the rectified flow approach based on optimal transport paths, as shown in Fig. 2. This paradigm has demonstrated superior generation quality and training efficiency compared to traditional diffusion models. Flow matching consists of forward and reverse processes [25].

In forward process, flow matching defines a probability path that transforms a simple noise distribution to the complex data distribution. Let $\mathbf{x}_1$ denote the clean scenario from the data distribution, and $\mathbf{x}_0 \sim \mathcal{N}(\mathbf{0}, \mathbf{I})$ denote the initial Gaussian noise. For a time step $t \in [0, 1]$, the conditional probability path $p(\mathbf{x}_t|\mathbf{x}_1)$ is defined as:

$$p(\mathbf{x}_t \mid \mathbf{x}_1) = \mathcal{N}(\mathbf{x}_t; t\mathbf{x}_1, (1-t)^2 \mathbf{I}) \quad (1)$$

where the state $\mathbf{x}_t$ evolves along an optimal transport path, which linearly interpolates between the noise and the data:

$$\mathbf{x}_t = t\mathbf{x}_1 + (1-t)\mathbf{x}_0 \quad (2)$$

The ground truth velocity field $\mathbf{v}_t$ governing this transition is constant and can be analytically derived as:

$$\mathbf{v}_t = \frac{d\mathbf{x}_t}{dt} = \mathbf{x}_1 - \mathbf{x}_0 \quad (3)$$

During reverse generation process, a neural network velocity estimator $\mathbf{v}_\theta$ is trained to approximate the ground truth velocity field. Given the current state $\mathbf{x}_t$, time $t$, and text embedding $\mathbf{c}$, the network outputs the predicted velocity vector $\mathbf{v}_{pred}$.

$$\mathbf{v}_{pred} = \mathbf{v}_\theta(\mathbf{x}_t, t, \mathbf{c}) \quad (4)$$

Flow matching models typically rely solely on the Mean Squared Error (MSE) in the time domain. However, relying exclusively on time-domain supervision often suffers from spectral bias, where the model prioritizes learning low-frequency components while neglecting high-frequency details. This is detrimental for scenario generation, where high-frequency fidelity is crucial for system simulation.

To mitigate this, an auxiliary spectral consistency loss $\mathcal{L}_{freq}$ is introduced. The predicted velocity $\mathbf{v}_{pred}$ and the target velocity $\mathbf{v}_t$ are transformed into the frequency domain using the Fast Fourier Transform (FFT). The frequency loss measures the discrepancy between the magnitudes of spectral components. Mean absolute error is used instead of MSE because spectral magnitudes span several orders of magnitude; MSE would amplify errors in large-amplitude components, biasing optimization toward them and causing instability:

$$\mathcal{L}_{freq} = \mathbb{E}_{\mathbf{v}_{pred},\mathbf{v}_t}\left[\|\mathcal{F}(\mathbf{v}_{pred}) - \mathcal{F}(\mathbf{v}_t)\|_1\right] \quad (5)$$

where $\mathcal{F}(\cdot)$ denotes the FFT operation and $\|\cdot\|_1$ represents the is the element-wise $\ell_1$ norm. While combining time-domain and frequency-domain losses is essential, simply minimizing a static weighted sum $\lambda$ is suboptimal:

$$\mathcal{L} = \underbrace{\mathbb{E}_{t,\mathbf{x}_0,\mathbf{x}_1}\left[\|\mathbf{v}_{pred} - \mathbf{v}_t\|_2^2\right]}_{\mathcal{L}_{time}} + \lambda \mathcal{L}_{freq} \quad (6)$$

The optimization landscapes of these two objectives often conflict: $\mathcal{L}_{time}$ prioritizes smoothness, whereas $\mathcal{L}_{freq}$ encourages high-frequency variance. To address this gradient conflict, the training is formulated as a multi-objective optimization problem, aiming to find a Pareto optimal solution where spectral fidelity is maximized without compromising temporal accuracy. The MGDA is employed to dynamically balance the gradients of the two objectives [26]. Let $\nabla_\theta \mathcal{L}_{time}$ and $\nabla_\theta \mathcal{L}_{freq}$ denote the gradients with respect to the shared parameters. A dynamic weight $\alpha \in [0, 1]$ is determined to form a combined gradient direction that decreases both losses:

$$\min_{\alpha \in [0,1]} \|\alpha \nabla_\theta \mathcal{L}_{time} + (1-\alpha)\nabla_\theta \mathcal{L}_{freq}\|_2^2 \quad (7)$$

$$\alpha = \text{clip}\left(\frac{(\nabla_\theta \mathcal{L}_{freq} - \nabla_\theta \mathcal{L}_{time})^\top \nabla_\theta \mathcal{L}_{freq}}{\|\nabla_\theta \mathcal{L}_{time} - \nabla_\theta \mathcal{L}_{freq}\|_2^2}, 0, 1\right) \quad (8)$$

where $\text{clip}(\cdot, 0, 1)$ constrains the weight to the valid range. This quadratic problem has an efficient analytical solution, allowing the model to automatically adjust its focus between global trends and local details throughout the training process, ensuring robust convergence to the Pareto front. The training process can be conducted in Algorithm 1.

During inference, starting from pure noise $\mathbf{x}_0 \sim \mathcal{N}(\mathbf{0}, \mathbf{I})$, the target scenario $\mathbf{x}_1$ is obtained by numerically solving the Ordinary Differential Equation (ODE) defined by the learned velocity field:

$$d\mathbf{x}_t = \mathbf{v}_\theta(\mathbf{x}_t, t, \mathbf{c})dt \quad (9)$$

This formulation allows for straight-line trajectories in the flow, enabling high-fidelity sampling with fewer steps compared to stochastic diffusion processes.

**Algorithm 1:** Train flow diffusion
**Input:** Dataset $\mathcal{D}=\{(\mathbf{x}_1^{(i)}, \mathbf{c}^{(i)})\}_{i=1}^N$, initialized model $\mathbf{v}_\theta$,
LLM $\Phi$, learning rate $\eta$
**while** not converged **do**
  Draw batch $\{(\mathbf{x}_1, \mathbf{c})\}$ from $\mathcal{D}$
  Sample noise $\mathbf{x}_0 \sim \mathcal{N}(\mathbf{0}, \mathbf{I})$ and time $t \sim \mathcal{U}(0, 1)$
  Get semantic embedding: $\mathbf{E}_{txt} \leftarrow \Phi(\mathbf{c})$
  Interpolate state $\mathbf{x}_t \leftarrow t\mathbf{x}_1 + (1-t)\mathbf{x}_0$ and target $\mathbf{v}_t = \mathbf{x}_1 - \mathbf{x}_0$
  Predict velocity: $\mathbf{v}_{pred} = \mathbf{v}_\theta(\mathbf{x}_t, t, \mathbf{c})$
  Compute gradients for Time($\mathbf{g}_t$) and Frequency($\mathbf{g}_f$)
  Compute Pareto-optimal weight $\alpha$
  Update parameters: $\theta \leftarrow \theta - \eta(\alpha \mathbf{g}_t + (1-\alpha)\mathbf{g}_f)$
**end while**

### B. Data Encoding

*1) Semantic encoding via LLM*

To enable the synthesis of time series from arbitrary open-domain descriptions, a robust semantic guidance mechanism driven by a pre-trained LLM is introduced. This mechanism leverages the extensive linguistic knowledge and reasoning capabilities encapsulated in modern open-source LLMs to extract rich contextual representations.

Formally, a specific tokenizer $\mathcal{T}$ is employed to transform the natural language instruction $\mathbf{c}$ into a sequence of discrete tokens $S$:

$$S = \mathcal{T}(\mathbf{c}) = \{w_1, w_2, \ldots, w_M\} \quad (10)$$

where $M$ represents the sequence length. Subsequently, a pre-trained LLM encoder, parameterized by $\Phi$, maps this token sequence into a high-dimensional continuous semantic space. The contextualized embedding $\mathbf{E}_{txt}$ is obtained from the hidden states of the final transformer layer:

$$\mathbf{E}_{txt} = \text{LLM}_\Phi(S) \in \mathbb{R}^{M \times D_{LLM}} \quad (11)$$

where $D_{LLM}$ denotes the hidden dimension of the LLM. This embedding $\mathbf{E}_{txt}$ encapsulates both local semantic details and global textual context. To ensure training stability, the pre-trained LLM is frozen during the optimization of the velocity predictor. The resulting embedding $\mathbf{E}_{txt}$ serves as the conditional input in subsequent stage, guiding the generation process without incurring the high computational cost of LLM fine-tuning.

*2) Time embedding*

The velocity field $\mathbf{v}_\theta(\mathbf{x}_t, t)$ varies significantly as the flow evolves from noise ($t=0$) to data ($t=1$). To capture this timestep-dependent evolution, the denoising network is conditioned on the diffusion timestep via a sinusoidal embedding that maps the scalar $t$ into a high-dimensional feature space. Specifically, the timestep $t$ is first scaled by a constant factor $s=100$ to adjust its sensitivity. The embedding $\mathbf{t}_{emb}$ is constructed by concatenating sine and cosine transformations across a spectrum of geometric frequencies. Formally, the embedding is defined as:

$$\mathbf{t}_{emb} = \left[\ldots, \sin\left(\frac{s \cdot t}{\omega_i}\right), \cos\left(\frac{s \cdot t}{\omega_i}\right), \ldots\right] \quad (12)$$

$$\omega_i = 10000^{\frac{2i}{d}} \quad (13)$$

This encoding ensures that the timestep representation remains continuous and differentiable, providing a multi-scale reference for the downstream blocks to distinguish different stages of the generative process.

### C. Text-oriented Temporal Denoising Network

The efficacy of the rectified flow framework critically depends on the capacity of the velocity estimator $\mathbf{v}_\theta$. A text-oriented temporal denoising network is proposed, which serves as a specialized architecture engineered to map the complex interplay between semantic conditions and continuous temporal dynamics. As illustrated in Fig. 3, this network integrates a hierarchical convolutional backbone with a cross-modal attention mechanism, ensuring that the generated scenarios exhibit both local temporal coherence and global semantic alignment.

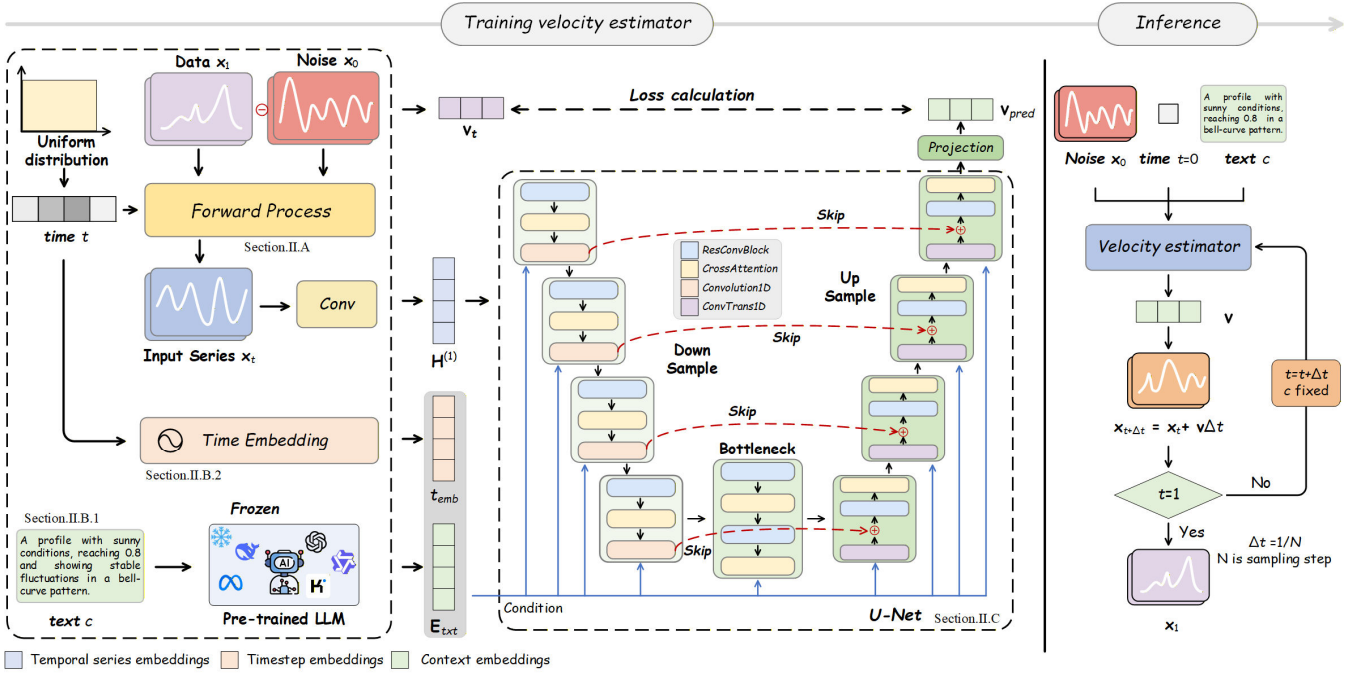

Fig. 3. Overview of the LFFD framework. The architecture employs a rectified flow matching conditioned on context embeddings from a frozen LLM. The U-Net-based velocity estimator integrates residual blocks for temporal modulation and cross-modal attention blocks to align discrete textual guidance with continuous temporal features. The model is trained to learn straight-line trajectories via a single-step objective, while inference is performed through multi-step sampling.

To accurately reconstruct scenarios that contain both low-frequency trends and high-frequency transients, a symmetric encoder-decoder U-Net topology with multi-scale feature processing is adopted.

- Encoding: The encoder captures hierarchical temporal patterns by progressively reducing the sequence length $L$ and expanding the channel dimension $C$. This is achieved through a stack of down-sampling blocks, each consisting of 1D convolutions with stride 2. This process compresses the input series into a compact latent representation, forcing the model to learn robust high-level features.
- Bottleneck: At the lowest resolution, a deep bottleneck stage processes the most abstract semantic features, capturing global temporal dependencies.
- Reconstruction: The decoder recovers the temporal resolution via transposed 1D convolutions. Crucially, to mitigate the information loss inherent in down-sampling, long-range skip connections are employed. These connections directly concatenate the high-resolution feature maps from the encoder to the corresponding decoder layers, creating an information highway that preserves fine-grained morphological details essential for high-fidelity generation.
- Projection: a terminal convolution layer projects the multi-channel features back to the original data dimension, yielding the predicted velocity $\mathbf{v}_{pred}$.

*1) Residual Block*

As shown in Fig. 4, residual convolutional block is the fundamental building unit. To inject the diffusion step condition without disrupting the feature sequence, a timestep-aware additive modulation mechanism is employed.

Within each block, the diffusion timestep embedding $\mathbf{t}_{emb}$ is projected into the feature space of the $l$-th layer via a layer specific linear transformation to generate a bias vector $\mathbf{t}_{bias}^{(l)}$ via a layer-specific linear transformation. This bias vector is then injected into the middle feature map $\mathbf{H}_{mid}^{(l)}$ via channel-wise addition. The residual block is formally defined as follows:

$$\mathbf{H}_{res}^{(l)} = \text{Conv1d}(\mathbf{H}^{(l)}) \tag{14}$$

$$\mathbf{H}_{mid}^{(l)} = \text{Conv1d}(\text{SiLU}(\text{GroupNorm}(\mathbf{H}_{res}^{(l)}))) \tag{15}$$

$$\mathbf{t}_{bias}^{(l)} = \text{Linear}(\mathbf{t}_{emb}) \tag{16}$$

$$\mathbf{H}_{mod}^{(l)} = \text{Conv1d}(\text{SiLU}(\text{GroupNorm}(\mathbf{H}_{mid}^{(l)} \oplus \mathbf{t}_{bias}^{(l)}))) \tag{17}$$

$$\mathbf{H}_{out}^{(l)} = \mathbf{H}_{res}^{(l)} + \mathbf{H}_{mod}^{(l)} \tag{18}$$

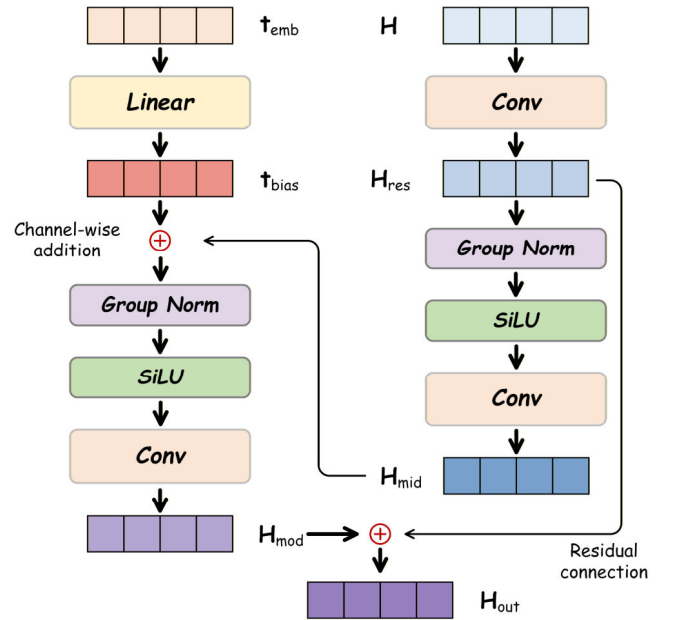

Fig. 4. Architecture of the residual block. It employs a timestep-aware additive modulation mechanism, where projected timestep embeddings are injected into feature maps via channel-wise addition to condition the denoising process without disrupting temporal features.

This additive modulation mechanism dynamically shifts the activation distribution of the feature maps, informing the network of the current noise level without compromising the temporal structure of the scenario features.

*2) Cross-Modal Attention Block*

The seamless integration of LLM embeddings $\mathbf{E}_{txt}$ into the generation process is realized through cross-attention layers strategically embedded at multiple resolutions of the network. This architectural design bridges the discrete semantic space and the continuous physical time domain. Mathematically, the intermediate temporal features $\mathbf{H}$ act as the query, while the text embeddings $\mathbf{E}_{txt}$ serve as both the key and value. The alignment is computed as:

$$\text{Attention} = \text{Softmax}\left(\frac{(\mathbf{H}\mathbf{W}_Q)(\mathbf{E}_{txt}\mathbf{W}_K)^T}{\sqrt{d_k}}\right)(\mathbf{E}_{txt}\mathbf{W}_V) \quad (19)$$

where $\mathbf{W}_Q$, $\mathbf{W}_K$, $\mathbf{W}_V$ are learnable projection matrices. The resulting attention matrix captures the relevance of each semantic token to specific temporal segments. This allows the network to attend to specific keywords and modulate the corresponding local intervals of the scenarios, ensuring that the synthesized result is not only realistic but also strictly adheres to the textual instructions.

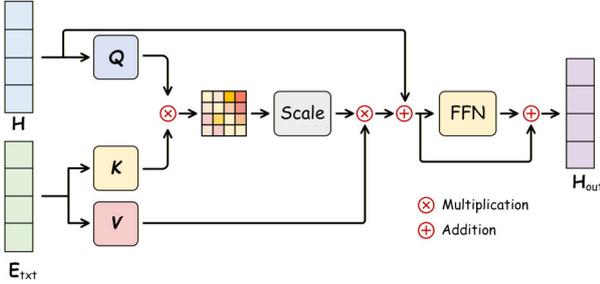

Fig. 5. Architecture of the cross-modal attention block. This module acts as a bridge between the discrete semantic space and continuous physical time domain by using temporal features as queries and text embeddings as keys and values to enforce strict semantic alignment.

As illustrated in Fig. 5, the cross-modal attention block incorporates residual connections: the attention output $\mathbf{H}_{att}$ is summed with the input features $\mathbf{H}$, then passed through group normalization and a residual Feed Forward Network (FNN). To enhance training stability, the block adopts a pre-normalization design.

$$\mathbf{H}_{att} = \text{Attention}(\text{LayerNorm}(\mathbf{H}), \mathbf{E}_{txt}) \quad (20)$$

$$\mathbf{H}_{attout} = \mathbf{H}_{att} + \mathbf{H} \quad (21)$$

$$\mathbf{H}_{output} = \text{FNN}(\text{LayerNorm}(\mathbf{H}_{attout})) + \mathbf{H}_{attout} \quad (22)$$

*D. LLM-Agents for Dataset Synthesis and Evaluation*

Realizing text-guided time series generation entails two primary obstacles: (1) the scarcity of high-quality paired datasets containing time series and their corresponding textual descriptions, and (2) the lack of metrics to quantify the semantic alignment between the generated scenarios and the text prompts. To address these challenges, a dual-agent framework driven by LLMs is proposed: The annotator agent for textual synthesis (Fig. 6(a)) and The judge agent for quality assurance (Fig. 6(b)).

Directly feeding raw time series into LLMs poses challenges due to tokenization artifacts, exemplified by the fragmentation of a single decimal value like "0.1" into separate tokens for "0", ".", and "1". This phenomenon hinders the model's ability to interpret numerical patterns effectively [27]. To mitigate this limitation, a statistical analysis module is utilized as a semantic bridge.

This module computes global metrics including volatility and extremes while performing segmented analysis across the dawn, morning, afternoon, and evening temporal windows to generate a structured statistical report. Finally, this report is fed into the LLM alongside the raw data to enhance numerical comprehension.

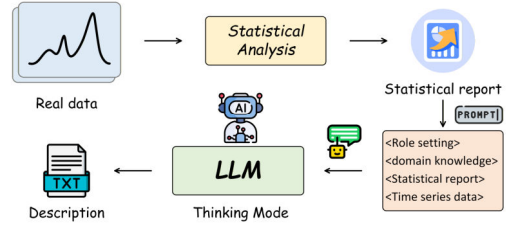
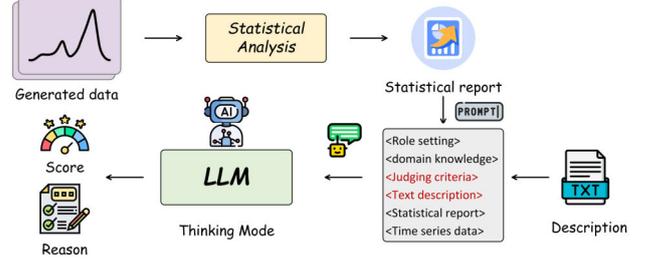

Fig. 6. The proposed LLM-driven dual-agent framework. (a) The annotator agent synthesizes high-quality textual descriptions for unlabeled scenarios using statistical reports and domain knowledge. (b) The judge agent acts as a semantic evaluator, scoring the alignment between the generated scenarios and the input text prompts to quantify controllability.

The annotator agent is engineered to automatically generate high-quality annotations for unlabeled real-world scenarios. This agent processes raw scenarios by utilizing a structured prompt composed of four integral elements, including a role setting that defines the persona of a data analyst, context-specific domain knowledge, the previously derived statistical report, and the numerical data itself. By leveraging the chain of thought reasoning capabilities of the LLM, the agent synthesizes these multi-dimensional inputs to produce a coherent textual description that accurately reflects the morphological and statistical characteristics of the underlying scenarios.

Complementing the scenarios annotation process, the judge agent serves to evaluate the semantic fidelity of the generated scenarios. Although its operational workflow parallels that of the annotator agent, the prompt engineering is adapted to include distinct evaluation criteria. The input is augmented to explicitly incorporate the original target text description which functions as the ground truth for comparison. Consequently, the agent is tasked with assessing the consistency between the statistical properties of the generated scenarios and the linguistic instructions. This process culminates in the output of a quantitative score along with a detailed justification, providing a robust and interpretable metric for quantifying the alignment between the generated scenarios and the guiding instructions.

## III. CASE STUDY

*A. Experimental Setups*

All experiments are implemented using the PyTorch framework and Python 3.10 on a high-performance computing cluster equipped with NVIDIA A100 GPUs with 80GB of memory. The velocity estimator $\mathbf{v}_\theta$ is instantiated as the proposed text-oriented temporal denoising network, following a 1D U-Net architecture with channel multipliers of {1, 2, 4, 8} and a base dimension of 64. To encode semantic guidance, a frozen pre-trained Large Language Model (e.g., Qwen-8B) is employed, projecting text inputs into 768-dimensional embeddings that are fused with temporal features via multi-resolution Cross-Attention blocks. The model is trained for 5,000 epochs with a global batch size of 256, using the AdamW optimizer and a OneCycleLR scheduler with a peak learning rate of $1 \times 10^{-4}$. For inference, the rectified flow formulation is adopted and solved by an Euler ODE solver with 50 discretization steps, ensuring a balance between generation quality and computational efficiency.

To evaluate the performance of the LFFD framework, comparisons are made against four representative generative models adapted for time-series synthesis: a VAE with a convolutional encoder-decoder to learn latent representations of scenarios; a deep convolutional GAN trained via the classic minimax adversarial game; Wasserstein GAN (WGAN), which stabilizes training and alleviates mode collapse through Wasserstein distance and gradient constraints; and DDPM implemented with the same U-Net backbone as the LFFD framework but trained under the standard noise-prediction objective ($\varepsilon$-prediction). The DDPM requires 1,000 iterative sampling steps, serving as a baseline for both generation quality and inference speed.

*B. Dataset Description*

To evaluate the generalizability of the proposed LFFD framework, two representative power time-series datasets characterized by distinct temporal granularities and scales are utilized. The first is a large-scale Photovoltaic (PV) dataset comprising over 32,000 daily scenarios with a 5-minute resolution. The second is an electricity load dataset containing over 2,000 daily scenarios with a 15-minute resolution. These datasets were selected to represent diverse power system scenarios, varying in dataset size, sampling frequency, and fluctuation characteristics. Prior to training, Min-Max normalization was applied to the PV dataset. For the load dataset, given the significant magnitude disparity between subsets, independent normalization was performed for the residential and industrial scenarios, respectively. All data were scaled to the range of [0, 1] to ensure numerical stability and model convergence. Subsequently, the annotator agent is employed to generate corresponding textual prompts. Addressing the difficulty in quantitatively evaluating open-ended natural language, the agent is designed to extract structured metadata regarding key statistical features alongside prompt generation. For instance, within the PV dataset, the agent labels the volatility category, providing ground truth for quantitative evaluation.

*C. Ordered Semantic Space of LLM*

To examine whether the LLM constructs a structured semantic space rather than relying on textual memorization, a linear probing analysis is conducted based on the linear representation hypothesis, as illustrated in Fig. 7. This hypothesis postulates that if a model effectively encodes specific concepts, such as load magnitude or environmental conditions, these concepts should be linearly decodable from its latent representations. Using inputs consisting of unstructured natural language descriptions, embeddings are extracted from the model's final hidden layer and mean-pooling is applied to obtain fixed-size vector representations. Subsequently lightweight linear probes are trained, employing linear regression for continuous variables and logistic regression for categorical attributes, to map these frozen embeddings to ground truth labels across weather, peak, volatility, and shape dimensions.

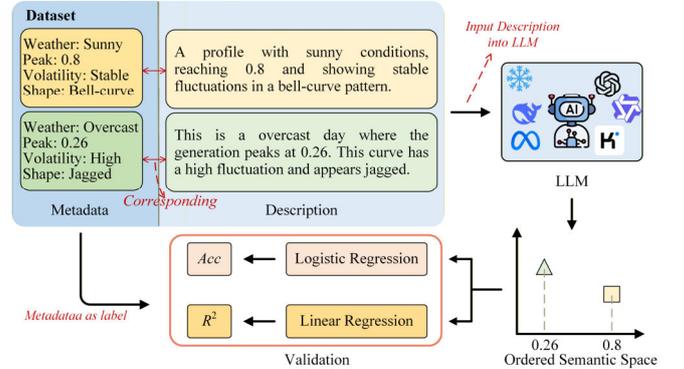

Fig. 7. Schematic of the composition of the dataset and the semantic space validation methodology. Metadata serves as ground-truth labels. The framework assesses whether the LLM's embeddings encode properties by measuring the decodability of these labels using lightweight linear probes based on the linear representation hypothesis.

The results in Fig. 8 indicate a highly structured latent space, as the linear probes achieved high performance across all tested dimensions. For the continuous "Peak" attribute, the linear regression analysis yielded a coefficient of determination ($R^2$) of 1.00, suggesting that physical magnitude is encoded as a precise vector direction within the high-dimensional space. Similarly, the probes achieved 100% classification accuracy for categorical attributes, indicating that the semantic manifolds for different classes are linearly separable. semantic nuance; the cluster for "Sunny" is positioned adjacent to "Sunny with Clouds" (S+C), reflecting their conceptual similarity, whereas the adverse weather conditions of "Rainy," "Cloudy," and "Stormy" are collectively situated on the opposing side of the projection space. These findings confirm that the LLM maps unstructured prompts onto a geometrically organized manifold where physical properties and semantic concepts are encoded orthogonally, verifying the model's capability to serve as a reliable semantic encoder for downstream power system tasks.

*D. Statistical Evaluation*

The primary objective of generative modeling is to generate scenarios that adhere to the inherent statistical distributions and physical laws of the domain. Conventional point-wise prediction metrics often fail to accurately characterize the generative capabilities, as they overlook the stochastic nature and morphological fidelity of scenarios. To assess the performance of the generative models and the quality of the generated scenarios, Kullback-Leibler divergence (KL), Maximum Mean Discrepancy (MMD), Fréchet Distance (FD), Dynamic Time Warping (DTW) and Power Spectral Density Distance (PSDD) are employed.

KL divergence measures the asymmetric difference between the probability distribution of the generated data $Q(x)$ and that of the real data $P(x)$. It is particularly effective for assessing whether the model captures the marginal distributions of the scenarios. The continuous time-series values are discretized to estimate probability densities, and the divergence is calculated as:

$$D_{KL}(P \| Q) = \sum_x P(x) \log \frac{P(x)}{Q(x)} \quad (23)$$

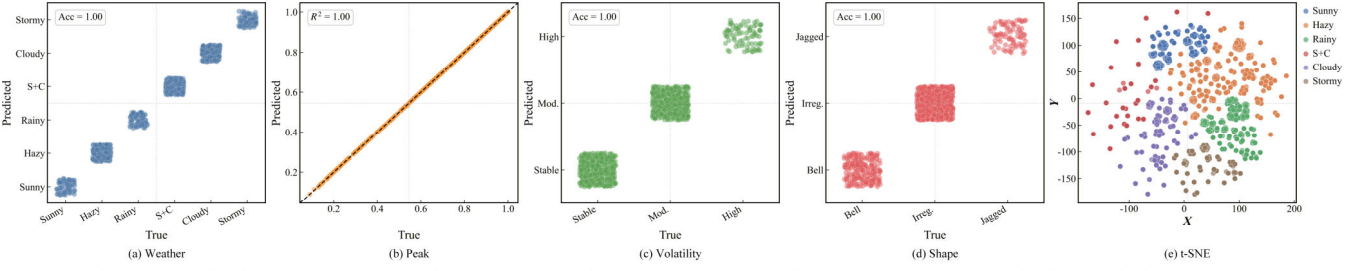

Fig. 8. Validation of the linear representation hypothesis. (a)–(d) Linear probing results for weather, peak, volatility, and shape, showing perfect classification accuracy and regression fit. (e) t-SNE visualization of the embedding manifold, highlighting the topological preservation of semantic relationships among different weather conditions. Note: S+C denotes sunny with clouds.

MMD serves as a kernel-based metric to quantify the distance between the true distribution $P$ and the generated distribution $Q$. By mapping samples into a reproducing kernel Hilbert space, MMD measures the distance between the mean embeddings of the two distributions, where a lower value indicates statistical proximity to the ground truth. Given real samples $X$ and generated samples $Y$, the squared MMD is empirically estimated as:

$$\text{MMD}^2(X,Y) = \frac{1}{m^2}\sum_{i,j}k(x_i,x_j) + \frac{1}{n^2}\sum_{i,j}k(y_i,y_j) \\ - \frac{2}{mn}\sum_{i,j}k(x_i,y_j) \quad (24)$$

where $k(\cdot,\cdot)$ represents the kernel function.

FD evaluates both the realism and diversity of generated samples by calculating the distance between feature distributions extracted by a pre-trained encoder. Assuming the feature vectors follow a multidimensional Gaussian distribution, FD is defined as:

$$\text{FD} = \| \mu_r - \mu_g \|^2 + \text{Tr}(\Sigma_r + \Sigma_g - 2(\Sigma_r \Sigma_g)^{1/2}) \quad (25)$$

where $\mu$ is the mean of feature vectors, $\Sigma$ is the covariance matrix and $\text{Tr}(\cdot)$ is the trace of a matrix. A lower FD score corresponds to higher generation quality and diversity.

DTW is used to evaluate shape fidelity by finding an optimal alignment between two temporal sequences, making it robust to temporal shifts and speed variations commonly found in scenarios. For two sequences $X$ and $Y$, DTW computes the minimum cumulative distance along an optimal warping path $W$:

$$\text{DTW}(X,Y) = \min_W \sum_{(i,j) \in W} d(x_i, y_j) \quad (26)$$

PSDD measures the fidelity of frequency-domain characteristics in generated scenarios, assessing how well periodic patterns and high-frequency variations are preserved. It computes the squared Euclidean distance between the log-average power spectra of real and generated data in the frequency domain:

$$D_{PSDD} = \left\| \log\left(\mathbb{E}[P_{real}(f)]\right) - \log\left(\mathbb{E}[P_{gen}(f)]\right) \right\|_2^2 \quad (27)$$

where $f$ represents the frequency components derived via Discrete Fourier Transform (DFT), and $\mathbb{E}[\cdot]$ denotes the average power spectrum across the dataset. A lower PSD distance indicates that the generated scenarios successfully retains the spectral energy distribution of the original scenarios, avoiding high-frequency information loss.

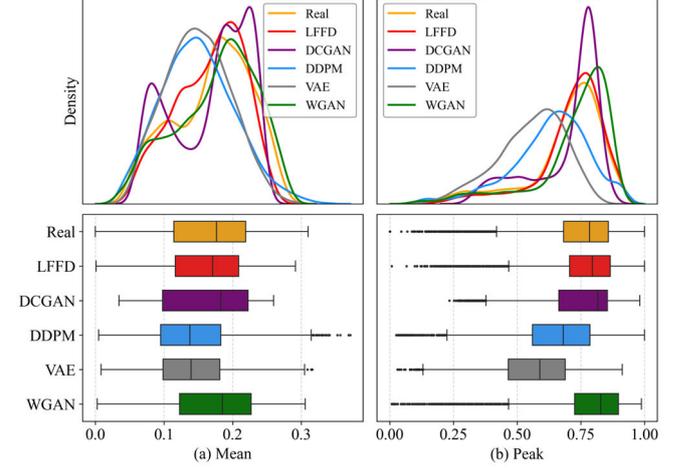

Fig. 9. (a) Mean PV distribution curves and boxplots, and (b) Peak PV distributions and boxplots.

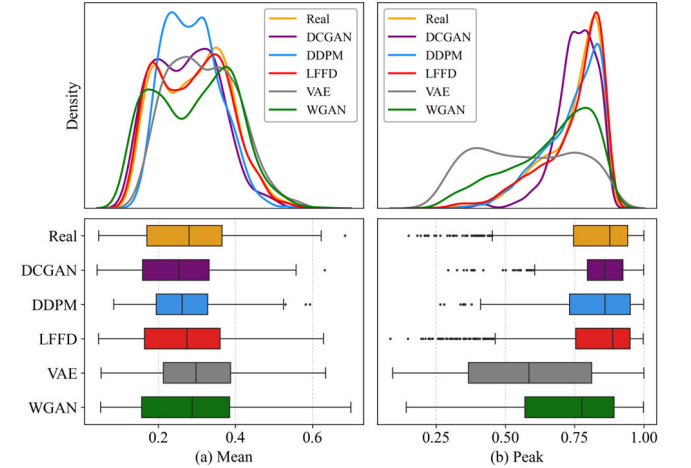

Fig. 10. (a) Mean load distribution curves and boxplots, and (b) Peak load distributions and boxplots.

To comprehensively evaluate the generative fidelity, the probability density distributions and statistical metrics of the synthesized samples are compared against the ground truth for both PV and load datasets. Figs. 9 and 10 illustrate the Probability Density Function (PDF) curves and feature boxplots, while Table I presents a rigorous quantitative evaluation using five metrics covering distributional alignment, sample quality, and frequency consistency. Visual inspection of the PDF curves reveals that real-world PV and load scenarios exhibit complex non-Gaussian and multi-modal characteristics. The proposed LFFD method demonstrates superior alignment with the ground truth, accurately reproducing both central tendencies and long-tail

TABLE I
MODEL PERFORMANCES ON DIFFERENT DATASETS

| Dataset | PV | | | | | Load | | | | |
|---|---|---|---|---|---|---|---|---|---|---|
| Model | KL | MMD | FD | DTW | PSDD | KL | MMD | FD | DTW | PSDD |
| VAE | <u>0.1163</u> | <u>0.2172</u> | 4.9652 | 27.0671 | 15.2839 | 0.3911 | 0.3471 | 1.7372 | 17.9072 | 14.8697 |
| DCGAN | 0.3922 | 0.9890 | 57.1187 | 28.9173 | 1.9987 | 0.1274 | 0.2316 | 0.9976 | 17.1228 | 5.5094 |
| WGAN | 0.2538 | 0.6434 | 20.7024 | 42.1868 | 1.9483 | 0.0484 | 0.1195 | 0.8288 | 16.1519 | 3.4473 |
| DDPM | 0.0836 | 0.2231 | <u>2.8568</u> | <u>24.0835</u> | <u>0.3957</u> | <u>0.1707</u> | <u>0.0805</u> | <u>0.4647</u> | <u>15.0642</u> | <u>0.1208</u> |
| LFFD | **0.0291** | **0.0905** | **0.9698** | **12.8813** | **0.0171** | **0.0373** | **0.0001** | **0.0073** | **14.8468** | **0.0011** |

*Bold term indicates the best performance, while underlining term represents second best.

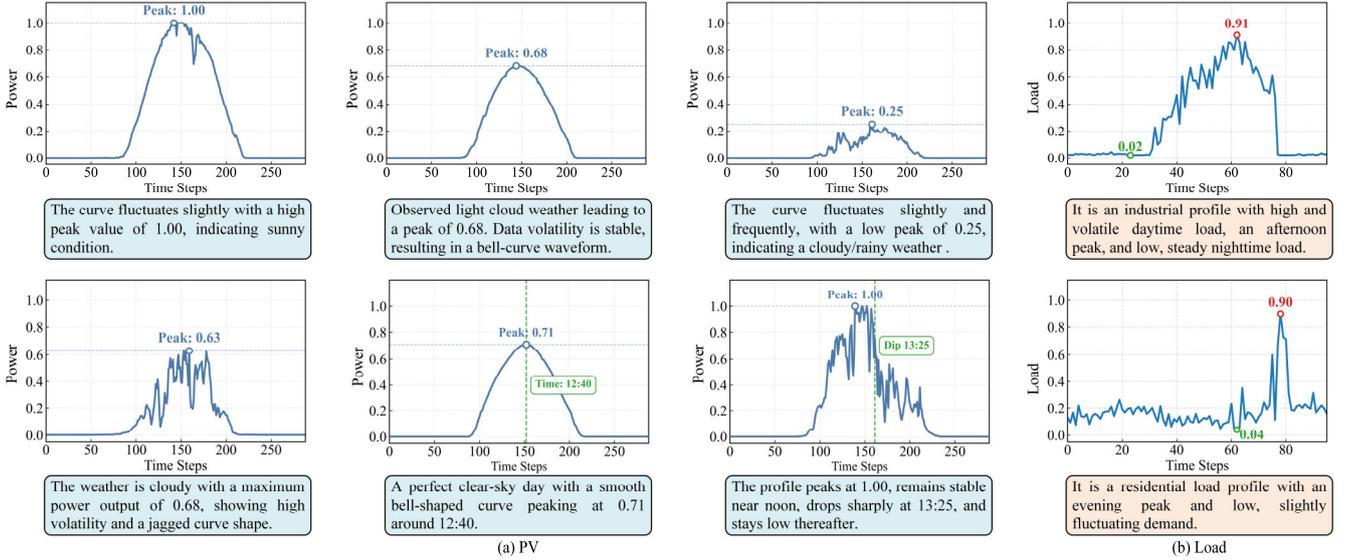

Fig. 11. Visualization of text-conditional scenario generation using LFFD. (a) PV generation scenarios capturing diverse weather conditions and fine-grained temporal features. (b) Load profiles distinguishing between industrial and residential consumption patterns. Annotated markers demonstrate the precise alignment between the generated waveforms and the input textual instructions.

fluctuations. In contrast, baseline models exhibit distinct limitations: VAE yields over-smoothed distributions that lack high-frequency details, whereas DCGAN and WGAN display noticeable deviations in peak positions and probability densities, indicative of mode collapse or distribution shifts.

Boxplot analysis further confirms LFFD's advantage in preserving scenarios dispersion and extremum properties. Regarding mean and peak statistics, LFFD maintains high consistency with real scenarios across medians, interquartile ranges, and whisker extents. While models like DDPM and WGAN often manifest dynamic range compression, LFFD successfully preserves the temporal correlations and statistical diversity of the original scenarios, attributed to its specialized temporal guidance mechanism.

As summarized in Table I, quantitative results corroborate these observations. LFFD outperforms baselines across all metrics, demonstrating a significant advancement over existing paradigms. High FD and DTW scores in VAE reflect its tendency toward over-smoothing, while elevated KL and MMD scores in adversarial models highlight inherent training instabilities. Although standard DDPM serves as a robust baseline, LFFD surpasses it by a substantial margin.

Regarding distributional alignment, LFFD achieves the lowest KL and MMD scores; notably, it reduces KL divergence on the PV dataset from 0.0836 to 0.0291. This indicates that our flow matching objective, combined with semantic guidance, captures complex marginal distributions more accurately than standard diffusion processes. Most critically, LFFD effectively addresses the spectral bias common in deep generative models. The PSDD is reduced by an order of magnitude compared to the strongest baseline. This validates the efficacy of the spectral-aware objective in retaining essential high-frequency transients. Consequently, LFFD achieves holistic superiority by balancing temporal dynamics with spectral accuracy and distributional fidelity.

*E. Text-Guidance Effectiveness*

To evaluate the efficacy of textual guidance in steering the generative process, a qualitative inspection of the synthesized samples is conducted as depicted in Fig. 11. The visual results demonstrate a precise alignment between the fine-grained semantic constraints provided in the prompts and the morphological characteristics of the generated scenarios. In the PV scenarios shown in Fig. 11(a), the model exhibits high fidelity to both meteorological conditions and numerical targets. When prompts specify sunny conditions or light clouds, the model generates smooth, standard bell-curve trajectories that adhere strictly to target peak magnitudes. Conversely, descriptions of adverse weather trigger the generation of jagged, non-stationary waveforms with frequent fluctuations, effectively capturing the intermittency of renewable generation. Crucially, the model demonstrates exceptional temporal localization capabilities; as evidenced in the bottom-row examples, it accurately renders specific time-dependent instructions, such as a peak occurring precisely at "12:40" or a sudden power "dip at 13:25," verifying its ability to enforce fine-grained temporal constraints.

Similarly, for the load dataset presented in Fig. 11(b), the model successfully disentangles complex consumption patterns based on user-type descriptions. A prompt describing an industrial scenario yields a curve characterized by high, volatile daytime consumption, whereas a residential scenario results in a distinctive evening peak with steady nighttime demand. These observations confirm that the model dynamically reconstructs the temporal geometry of the scenarios according to specific semantic instructions regarding shape, magnitude, volatility, and precise timing events.

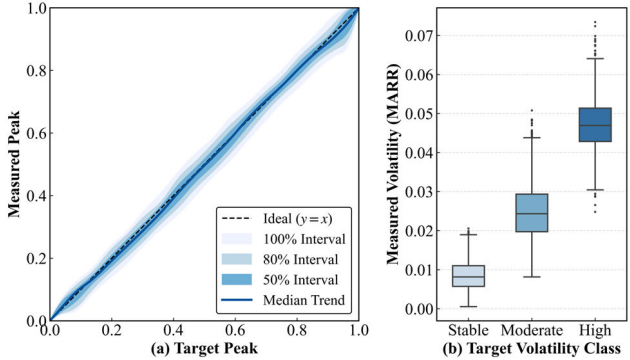

Fig. 12. Evaluation of attribute controllability. (a) Correlation between target and measured peak values, exhibiting strong alignment with the ideal trend. (b) Distribution of MARR across Stable, Moderate, and High target classes.

Complementing these visual observations, a quantitative assessment is performed to verify the statistical fidelity of the text-conditional generation with a specific focus on peak magnitude and volatility control. As illustrated in Fig. 12(a), the correlation between the target peak specified in the textual prompts and the actual measured peak of the generated samples is analyzed. The results exhibit a strong linear relationship where the median trend adheres closely to the ideal identity line ($y=x$) and exhibits narrow confidence intervals, thereby confirming the numerical sensitivity of the model to magnitude instructions. Subsequently, to evaluate the volatility control, the Mean Absolute Ramp Rate (MARR), a critical metric in power system analysis for capturing the instantaneous variability, is adopted. Given a generated power sequence $P = \{p_t\}_{t=1}^{T}$, the MARR is calculated as the expected value of the absolute first-order differences:

$$\text{MARR} = \frac{1}{T-1} \sum_{t=1}^{T-1} |p_{t+1} - p_t| \tag{28}$$

where $p_{t+1} - p_t$ represents the ramp event between adjacent time steps. This formulation explicitly quantifies the non-stationarity and the intensity of high-frequency components within the scenarios. Fig. 12(b) presents the distribution of MARR across the three textual volatility levels of stable, moderate, and high. The resulting box plots reveal a strict monotonic increase in measured volatility corresponding to the textual intensity. The clear separation between the interquartile ranges of the stable and high categories indicates that the model has effectively learned to manipulate local fluctuation characteristics based on discrete semantic categories, ensuring that the generated scenarios preserves both global trends and local statistical properties.

Furthermore, the alignment between textual instructions and the generated scenarios is quantified using the Mean Judge Agent Score (MJAS), as detailed in Section II.D. MJAS serves as a credible and comprehensive metric for semantic evaluation, designed to emulate domain experts' judgments regarding the alignment quality between generated scenarios and their corresponding textual descriptions. It is formally defined as follows:

$$\text{MJAS} = \sum \text{Agent}(x, T) \tag{29}$$

where $x$ is the generated time series and $T$ represents the corresponding text description. The evaluation yielded high MJAS values of 4.80 for the PV dataset and 4.46 for the load dataset relative to a maximum score of 5, confirming robust alignment between textual instructions and generated scenarios. The superior performance observed in the PV domain is attributed to its stronger physical determinism and larger data scale, which facilitate a more explicit mapping from meteorological descriptions to time-series patterns compared to the complex and stochastic nature of load consumption.

### F. Inference Efficiency Analysis

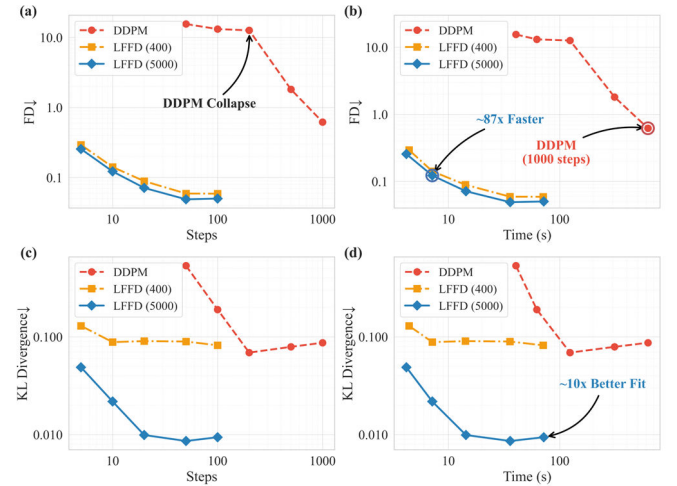

Fig. 13. Comparison of inference efficiency and generation quality between the DDPM baseline and the proposed LFFD method (at 400 and 5000 training epochs). (a) FD versus steps, illustrating the robustness of LFFD at low step counts compared to the quality collapse of DDPM. (b) FD versus inference time (seconds), highlighting the 87× speedup achieved by LFFD while maintaining superior quality. (c) KL Divergence versus steps. (d) KL Divergence versus inference time, demonstrating an order-of-magnitude improvement in distributional fit by the converged LFFD model. Both axes are plotted on a logarithmic scale.

To evaluate the deployment feasibility, a comparative analysis is conducted between the baseline DDPM and the proposed LFFD model on a PV test set of 3,200 samples. The evaluation focuses on the trade-off between generation quality and computational latency, as illustrated in Fig. 13. Results indicate that the LFFD model outperforms the DDPM baseline in terms of inference efficiency. While the standard DDPM requires 1,000 denoising steps to achieve its optimal FD of 0.62, incurring a computational cost of 621.68 seconds per batch, the proposed LFFD achieves a lower FD of 0.12 using 10 sampling steps, requiring 7.16 seconds. This corresponds to an approximate 87× speedup in wall-clock time and a 5-fold reduction in the FD metric, establishing a more favorable Pareto frontier compared to the baseline. Furthermore, the stability of the proposed architecture is evident under limited sampling steps. As shown in Fig. 13(a), the DDPM baseline exhibits significant performance degradation when sampling steps are reduced below 200, with FD scores increasing above 12.0. In contrast, attributed to the straightness of the learned ODE trajectories, LFFD maintains high fidelity even in the few-step regime; notably, at 5 sampling steps, LFFD achieves an FD of 0.26, surpassing the optimal performance of the DDPM baseline (FD 0.62 at 1,000 steps).

Beyond geometric fidelity, the statistical alignment of the generated scenarios is also evaluated using KL Divergence. The LFFD model trained for 5000 epochs achieves a KL divergence of 0.009, representing an order of magnitude improvement over the best result achieved by DDPM (0.087). This indicates that the proposed method captures the marginal distribution of the underlying physical process with significantly higher precision. Additionally, comparing LFFD-400 and LFFD-5000 reveals that while early-stage training is sufficient for capturing general waveform features, extended training is crucial for fine-grained distributional alignment, ensuring the generated scenarios adheres to the statistical laws.

## IV. Conclusion

This paper proposes LFFD, a framework that reconciles semantic interpretability with generative efficiency in power system analysis. First, by linearizing the inference trajectory through rectified flow matching, the model achieves an 87× speedup over traditional diffusion baselines. Second, the frequency-aware optimization effectively mitigates spectral bias, reducing frequency domain error by an order of magnitude and ensuring the accurate reconstruction of high-frequency transients. Third, integrating large language models enables the structured encoding of natural language, facilitating flexible, on-demand scenario generation beyond rigid fixed-label conditioning. Additionally, the developed dual-agent pipeline standardizes the synthesis of text-data pairs, addressing the scarcity of paired benchmarks. Consequently, LFFD provides a computationally efficient solution for generating high-fidelity, customized power system scenarios. Future work will focus on further scaling the framework to larger and more diverse power system dataset.